# Superconductivity of non- stoichiometric intermetallic compound $NbB_2$

Monika Mudgel[a,b], V.P. S. Awana[a, *], G. L. Bhalla[b] and H. Kishan[a]

[a]National Physical Laboratory, Dr. K.S. Krishnan Marg, New Delhi-110012, India

[b]Deptartment of Physics and Astrophysics, Delhi University, New Delhi-110007, India

## Abstract

We report the synthesis, magnetic susceptibility and crystal structure analysis for $NbB_{2+x}$ (x = 0.0 to 1.0) samples. The study facilitates in finding a correlation among the lattice parameters, chemical composition and the superconducting transition temperature $T_c$. Rietveld analysis is done on the X- ray diffraction patterns of all synthesized samples to determine the lattice parameters. The *a* parameter decreases slightly and has a random variation with increasing x, while *c* parameter increases from 3.26 for pure $NbB_2$ to 3.32 for x=0.4 i.e. $NbB_{2.4}$. With higher Boron content (x>0.4) the *c* parameter decreases slightly. The stretching of lattice in *c* direction induces superconductivity in the non-stoichiometric niobium boride. Pure $NbB_2$ is non-superconductor while the other $NbB_{2+x}$ (x>0.0) samples show diamagnetic signal in the temperature range 8.9-11K. Magnetization measurements (*M-H*) at a fixed temperature of 5K are also carried out in both increasing and decreasing directions of field. The estimated lower and upper critical fields ($H_{c1}$ & $H_{c2}$) as viewed from *M-H* plots are around 590 and 2000Oe respectively for $NbB_{2.6}$ samples. In our case, superconductivity is achieved in $NbB_2$ by varying the Nb/B ratios, rather than changing the processing conditions as reported by others.

*Corresponding Author:
Dr. V.P.S. Awana
Room 109,
National Physical Laboratory, Dr. K.S. Krishnan marg, New Delhi-110012, India
Fax No. 0091-11-25626938: Phone no. 0091-11-25748709
e-mail-awana@mail.nplindia.ernet.in: www.freewebs.com/vpsawana/

## Introduction

Studies on various diborides were greatly enhanced by the discovery of superconductivity in $MgB_2$ with a remarkably high transition temperature of 39 K[1]. $MgB_2$ is an intermetallic binary compound with P6/mmm crystal structure. The lighter Mg and B atoms contribute towards it's high $T_c$. So, the other $AlB_2$ type diborides were studied fundamentally and practically to search for high $T_c$. Various controversial reports exist on the superconductivity and value of $T_c$ for different diborides. For example, $ZrB_2$ is reported to have a $T_c$ of 5.5K by Gasprov et al [2], whereas Leyrovska and Leyrovski [3] report no transition. Similarly, Gasprov et al and others [2-5] have reported no observation of superconductivity in $TaB_2$ while Kackzorowski et al [6] report a transition temperature of 9.5K. The results for $NbB_2$ are even more diverse. Gasprov et al [2] Kackzorowski et al [6] reporting no superconductivity while many others [3, 6-10] report different values of transition temperature in the range 0.62 to 9.2K.

Moreover, synthesis of these diborides requires critical conditions of high pressure or arc melting etc. [5,11]. Avoiding these complexities we hereby report the synthesis of non- stoichiometric $NbB_2$ samples by simple argon annealing method at ambient pressure. The compositional dependence of structural and superconducting parameters like $T_c$ is studied systematically. The role of stretched $c$ parameter with the increased Boron content on the superconductivity of $NbB_2$ is described in the current communication.

## Experimental

Polycrystalline bulk samples of $NbB_{2+x}$ were synthesized by solid-state reaction route. The commercial $NbB_2$ and Boron powders were mixed in stoichiometric ratio according to the desired composition by continuous grinding. The well ground mixtures were palletized and encapsulated in iron tubes followed by sintering in a tubular furnace at 1100$^{o}$C in Argon flow for 20h. The ramp rate during heating was maintained to be 10$^{o}$/min. Then the samples were directly quenched to liquid nitrogen temperature. The phase formation was checked by X-ray diffraction patterns done on Rigaku-Miniflex-II at room temperature. Rietveld analysis was done by Fullprof program-2007 so as to obtain lattice parameters. Magnetic susceptibility measurements were carried out on a SQUID magnetometer (*MPMS-XL*).

## Results and Discussion

To understand the diversities of reported superconducting $T_c$, the structural phases of $NbB_2$ with different Nb/B ratios are realized. X-ray diffraction patterns of $NbB_{2+x}$ (x=0.0, 0.2, 0.4, 0.6, 0.8 & 1.0) are shown in Fig. 1. All the samples crystallize in P6/mmm, hexagonal structure. All characteristic peaks for the pure $NbB_2$ sample are indexed in the Figure. No extra impurity peak is noticed in any sample. Systematic shift is observed in (002) peak towards lower angle side with the increment in Boron content indicating the increase in *c* parameter. The enlarged view is shown in the inset of Fig.1. A single (002) peak is obtained up to $NbB_{2.8}$ i.e. for $Nb_{0.71}B_2$. Actually, the boron excess is incorporated into the phase creating metal vacancy in the lattice as discussed in various theoretical studies [12, 13]. The Boron plane is quite rigid and doesn't allow the extra boron to be incorporated at interstitial site. Hence the non-stoichiometry or Boron excess is accommodated by metal deficiency. For $NbB_{3.0}$, instead of a single (002) peak, a doublet is obtained indicating that both $NbB_2$ and $Nb_{1-x}B_2$ phase are present. It means that boron cannot be incorporated in the niobium boride lattice after 25-30 % limit of Niobium vacancy i.e. $Nb_{0.76}B_2$ to $Nb0_{.71}B_2$ .

Diffraction patterns are fitted using Rietveld analysis with the hexagonal $AlB_2$ structure model and space group P6/mmm (No. 191). Fig. 2(a) and 2(b) shows the Rietveld fitted diffraction patterns for $NbB_2$ and $NbB_{2.4}$. The differences between the experimental and calculated *XRD* patterns are very small. Lattice parameters are calculated for all samples by Rietveld analysis and are tabulated in Table 1. It is observed that *c* parameter increases continuously in the interval $0.0 \leq x \leq 0.4$ in $NbB_{2+x}$ samples. For pure $NbB_2$, $c$ = 3.26396(17)Å which increases sharply to 3.30509(18)Å and 3.32016(11)Å for $NbB_{2.2}$ & $NbB_{2.4}$ samples respectively. Beyond that *c* parameter changes slightly in a random way and hence has reached the saturation value. The occupancy factors are also calculated from the Rietveld analysis. As seen from Table 1, there is no considerable difference between the experimentally taken stochiometric ratios and the Rietveld determined values up to $NbB_{2.6}$ sample. After that the level of Boron incorporation in the lattice or the extent of metal vacancy creation seems to be saturated because the B/Nb ratio does not increase much after 2.6. The fact is also confirmed by the saturation of c parameter values after $NbB_{2.6}$ or $Nb_{0.76}B_2$. The extra boron forms $NbB_2$ phase along with the $Nb_{1-x}B_2$ phase as seen by a doublet in inset of Fig. 1.

To have a clear idea, the lattice parameters *a* & *c* and the ratio *c*/*a* are plotted in Fig. 3 with varying Boron content. The parameter *a* decreases slightly first and then does not change much. But the parameter *c* increases sharply with the boron content up to a certain level (x = 0.4) and then shows negligible up and downs. The *c*/*a* parameter changes exactly in the same way as the lattice parameter *c*. Thus, the lattice expands in *c*-direction with the boron excess. These structural changes are in confirmation with other reports [5, 14, 15].

In order to check the superconductivity of synthesized non-stoichiometric niobium boride samples, the magnetization measurements are carried out. The *M-T* plots are shown in figure 4 for $NbB_{2+x}$ samples. The pure $NbB_2$ sample doesn't give diamagnetic signal confirming that pure $NbB_2$ is non-superconductor. $NbB_{2.2}$ sample gives a very weak diamagnetic signal at a temperature of about 8.9K, which can only be seen in the enlarged view shown in the inset. The samples with higher boron content i.e. $NbB_{2+x}$ with $x \geq 0.4$ shows considerable diamagnetic signal at their respective transition temperature in the range 10-11K. The transition temperature is defined at the onset of diamagnetic signal. The inset of Fig. 4 shows the field cooled and zero field cooled magnetization curves for one of the composition $NbB_{2.4}$. This sample has a sufficient superconducting volume fraction.

In order to have a clear picture of variation of transition temperature with Boron content, the exact values of $T_c$'s are plotted in Fig. 5. The transition temperature increases continuously up to 11K for x=0.6 sample. Beyond that, it decreases slightly. The cell volume is also plotted which shows exactly the same behavior as $T_c$ with boron content. Thus, superconductivity is introduced in $NbB_2$ by increasing boron content or by creating Nb vacancy. The presence of vacancies in the Niobium sub-lattice of $NbB_2$ brings about considerable changes in the density of states in the near Fermi region and gives rise to a peak in the density of states [16]. The increase in the *DOS* (density of states) at fermi level corresponds to the increase in transition temperature with Boron excess.

Magnetization hysteresis loops (*M-H*) are shown in Fig. 6(a) for all synthesized samples in both the increasing and decreasing field directions at 5K. Pure $NbB_2$ sample doesn't show any negative moment, rather a paramagnetic signal is given which can be seen in the enlarged view in inset. $NbB_{2.2}$ sample gives weak negative moment with the field and possess a hysteresis in increasing and decreasing field directions. All other samples with greater boron content show considerable magnetic moments in opposite direction of field. All samples possess a magnetic hysteresis with respect to the direction of field. It is clear from the *M-H* plots that the non-stoichiometric niobium boride samples are Type-II superconductor. The similar behavior is reported earlier also for niobium deficient samples [17].

To estimate the values of lower critical field, $H_{c1}$ and upper critical field $H_{c2}$ for boron excess $NbB_2$ samples, the enlarged view of first quadrant of Fig. 6(a) is shown in Fig. 6(b). The $H_{c1}$ & $H_{c2}$ values are marked with arrows. The $H_{c1}$ is taken as the inversion point from where the diamagnetic moment starts decreasing or otherwise the field starts penetrating through the sample. $H_{c2}$ is taken as the field value at which the diamagnetic signal of the sample vanishes or otherwise the applied field completely penetrates through the sample. The lower critical field value $H_{c1}$ increases with increasing boron content and is observed to be maximum for the x=0.6 sample i.e about 592Oe. With further increase in Boron content, $H_{c1}$ value decreases to 479 Oe for x=1.0 i.e $NbB_3$ sample. The upper critical field values are almost same for $0.4\leq x\leq0.8$ samples of about 2000Oe while it is decreased to 1600Oe for $NbB_3$ sample.

## Conclusion

In summary, we report the structural and superconducting changes in non-stoichiometric niobium boride samples for the niobium deficient phases. The niobium vacancy cause the expanding of crystal lattice in *c*-direction thus increasing the *c*/*a* ratio and the cell volume. The upper limit to the metal vacancy creation is observed to be lie in range 25-30%. These structural changes are accompanied with the introduction of superconductivity. The transition temperature increases from 8.9-11 K with the increase of niobium vacancy. The *M-H* hysteresis loops confirm the type-II superconductivity in the metal deficient niobium boride samples. The lower critical field $H_{c1}$ increases with the increase in boron content up to x=0.6 sample with $H_{c1} \approx 592$Oe while decreases with further increment in boron content. The upper critical field value $H_{c2}$ is around 2000Oe for all super conducting samples except $NbB_3$ with $H_{c2} \approx 1600$Oe.

## Acknowledgement

The authors from *NPL* would like to thank Dr. Vikram Kumar (Director, *NPL*) for showing his keen interest in the present work. One of us Monika Mudgel would also thank *CSIR* for financial support by providing *JRF* fellowship. Authors would also like to thank Prof. E. Takayama-Muromachi from *NIMS* Japan for helping in carrying out the *SQUID* magnetization measurements.

## Figure Captions

Fig. 1 X- ray diffraction patterns of $NbB_{2+x}$ (x=0.0, 0.2, 0.4, 0.6, 0.8 & 1.0) samples in the angular range $20^o \leq 2\theta \leq 80^o$.

Fig. 2 Rietveld refined plots for (a) $NbB_2$ and (b) $NbB_{2.4}$ samples. X-ray experimental diagram (dots), calculated pattern (continuous line), difference (lower continuous line) and calculated Bragg position (vertical lines in middle).

Fig. 3 Variation of lattice parameters and *c/a* value with the increasing Boron content in non-stoichiometric Niobium Boride.

Fig. 4 Magnetization –Temperature measurements showing the transition temperature for $NbB_{2+x}$ samples with x=0.0, 0.2, 0.4, 0.6, 0.8 & 1.0. Lower inset shows the enlarged view for $NbB_2$ and $NbB_{2.2}$ samples. Upper inset is the FC and ZFC magnetization plots for $NbB_{2.4}$ sample.

Fig. 5 Cell Volume and superconducting transition temperature at different Boron contents.

Fig. 6(a) Magnetization hysteresis loops (*M-H*) for $NbB_{2+x}$ samples with x=0.0, 0.2, 0.4, 0.6, 0.8 & 1.0. Inset shows the enlarged view for $NbB_2$ and $NbB_{2.2}$ samples.

Fig. 6(b) Enlarged view of *M-H* loops for $NbB_{2+x}$ samples with x=0.4, 0.6, 0.8 & 1.0, the $H_{c1}$ & $H_{c2}$ values are marked by arrows.

Table 1: Lattice parameters, cell volume, *c*/*a* values and B/Nb ratios for $NbB_{2+x}$ samples with x = 0.0, 0.2, 0.4, 0.6, 0.8 & 1.0.

| x in $NbB_{2+x}$ | *a* (Å) | *c* (Å) | *Volume* (Å $^3$) | *c*/*a* | B/Nb (stoichiometric ratios ) | B/Nb (estimated from Rietveld Fits) |
|---|---|---|---|---|---|---|
| 0.0 | 3.11032(13) | 3.26396(17) | 27.345(2) | 1.049 | 2.0 | 2.038 |
| 0.2 | 3.10132(13) | 3.30509(18) | 27.531(2) | 1.066 | 2.2 | 2.184 |
| 0.4 | 3.10416(18) | 3.32016(11) | 27.706(1) | 1.069 | 2.4 | 2.405 |
| 0.6 | 3.10187(10) | 3.31951(14) | 27.660(2) | 1.070 | 2.6 | 2.626 |
| 0.8 | 3.10397(11) | 3.31718(15) | 27.678(2) | 1.069 | 2.8 | 2.614 |
| 1.0 | 3.10246(10) | 3.31961(11) | 27.670(1) | 1.070 | 3.0 | 2.674 |

Fig.1

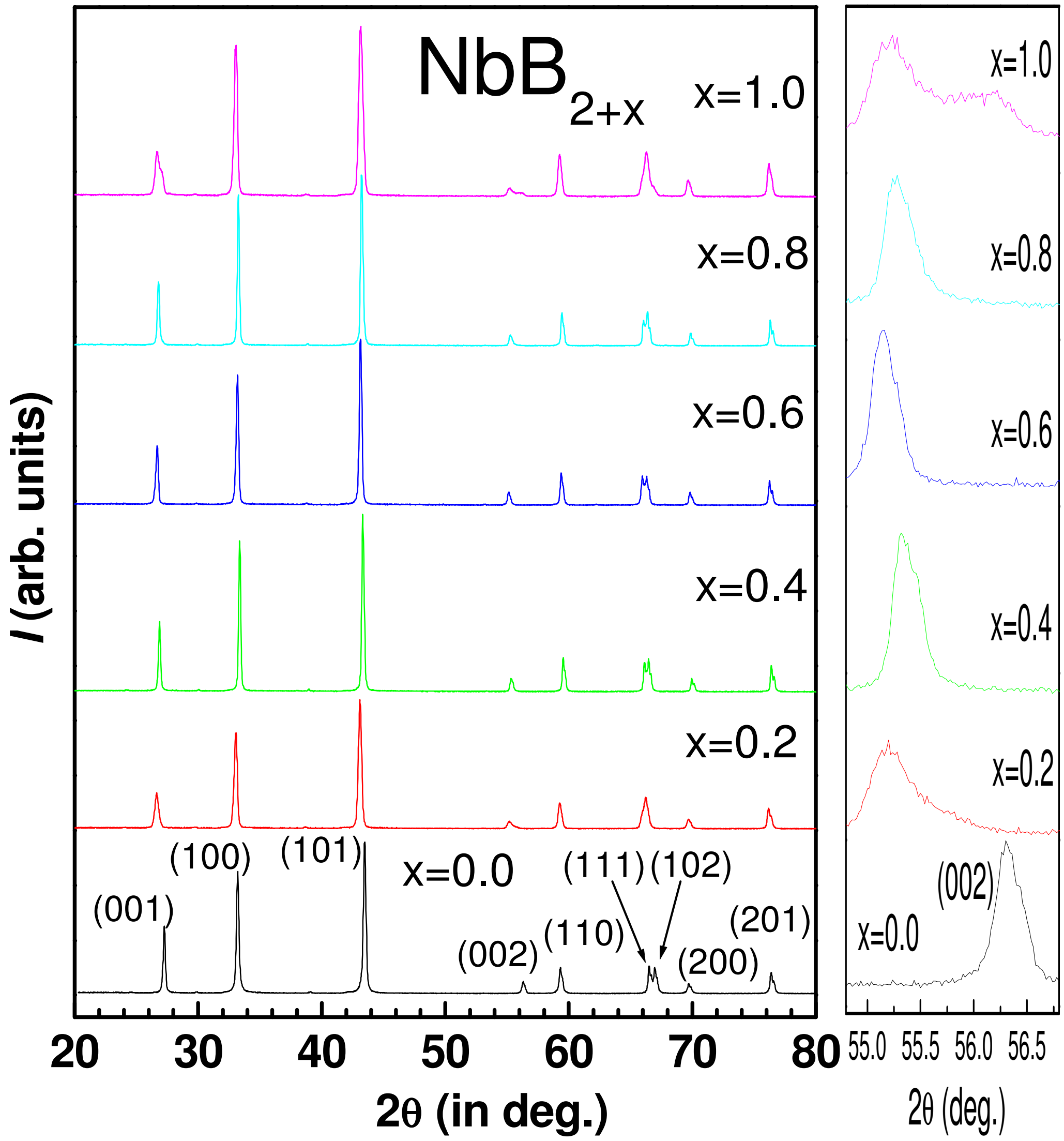

Fig.2 (a)

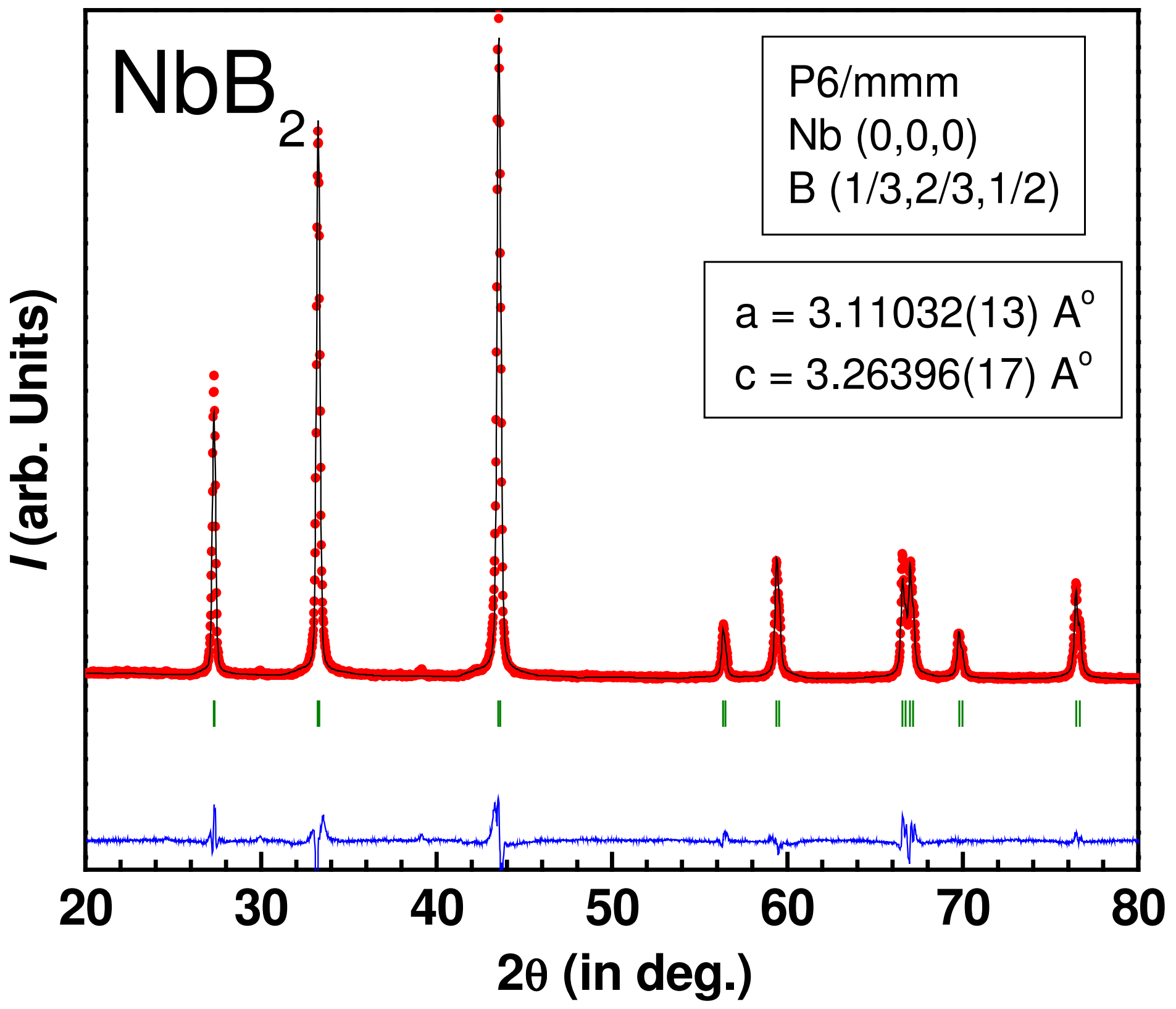
NbB2
P6/mmm
Nb (0,0,0)
B (1/3,2/3,1/2)
a = 3.11032(13) A°
c = 3.26396(17) A°
I (arb. Units)
20 30 40 50 60 70 80
2θ (in deg.)

Fig.2 (b)

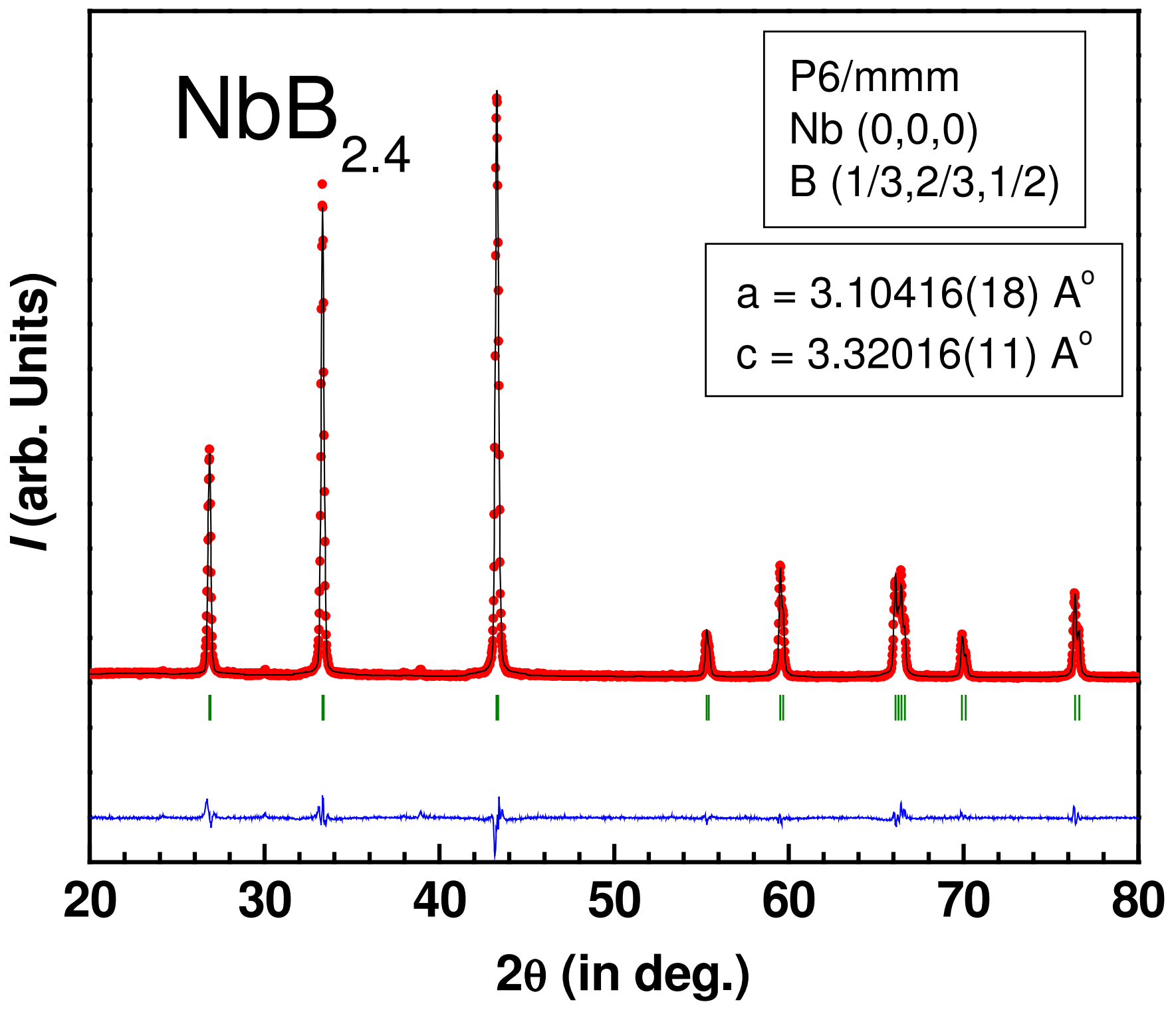
NbB2.4
P6/mmm
Nb (0,0,0)
B (1/3,2/3,1/2)
a = 3.10416(18) A°
c = 3.32016(11) A°
I (arb. Units)
20 30 40 50 60 70 80
2θ (in deg.)

Fig.3

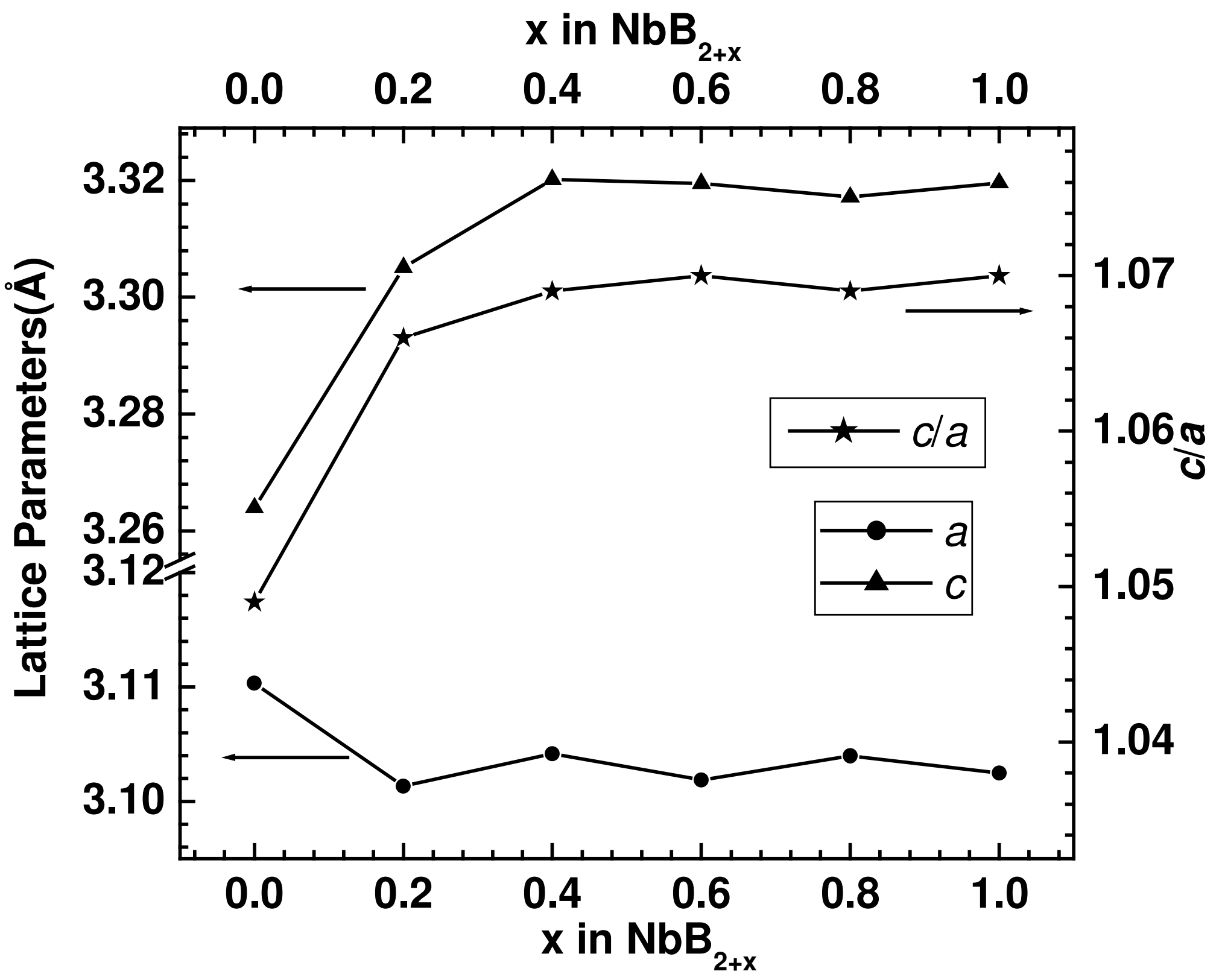

Fig.4

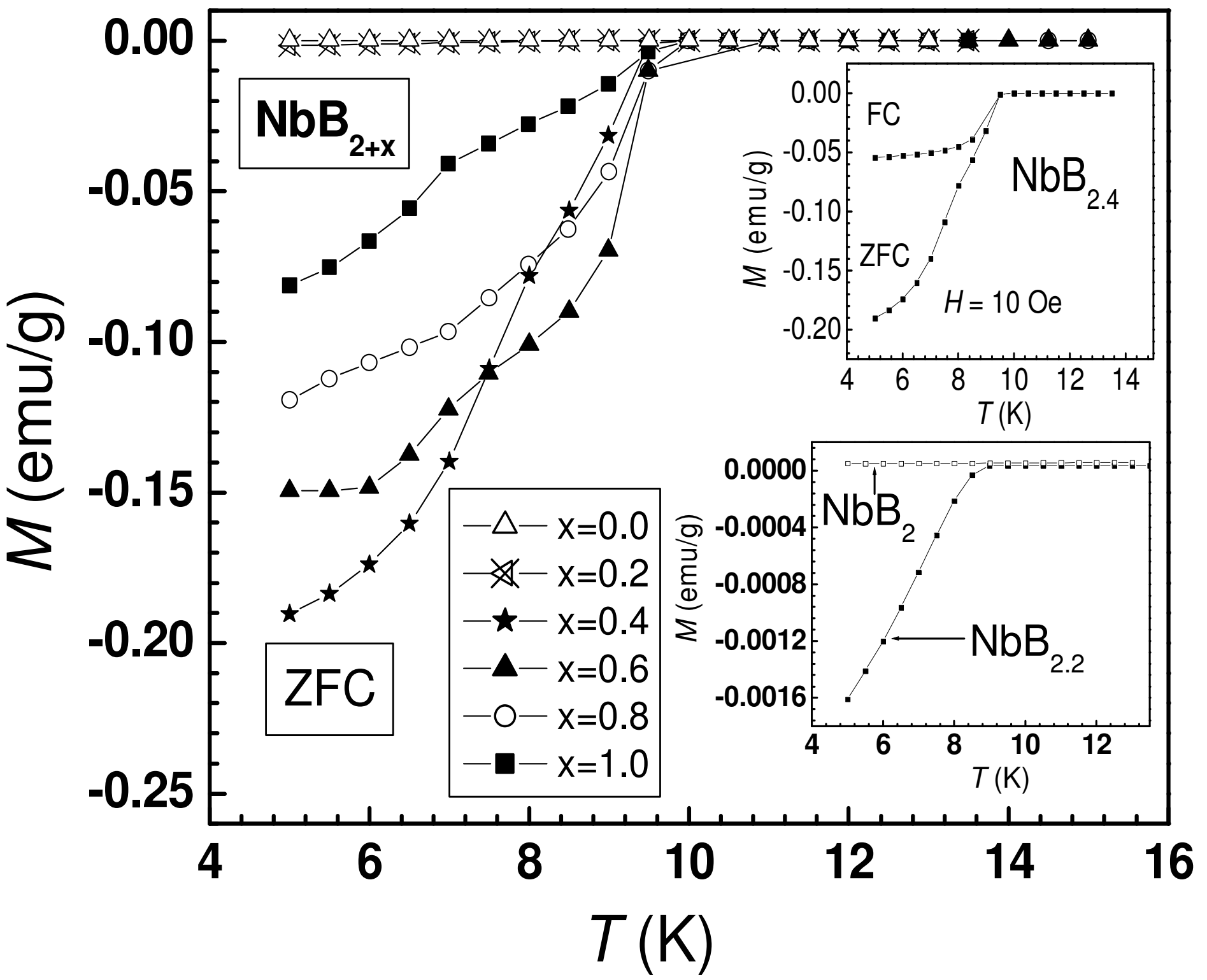

Fig. 5

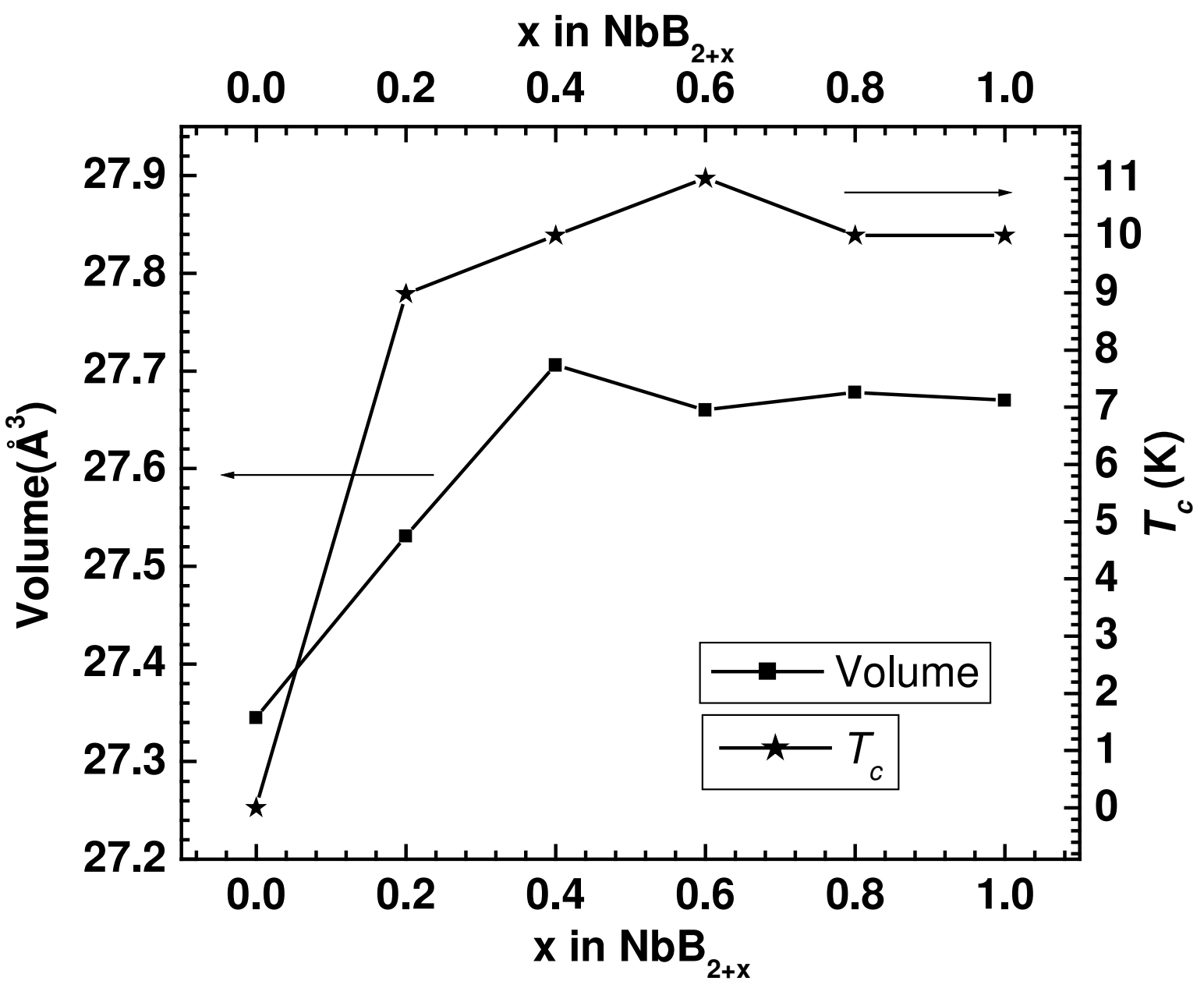

Fig.6(a)

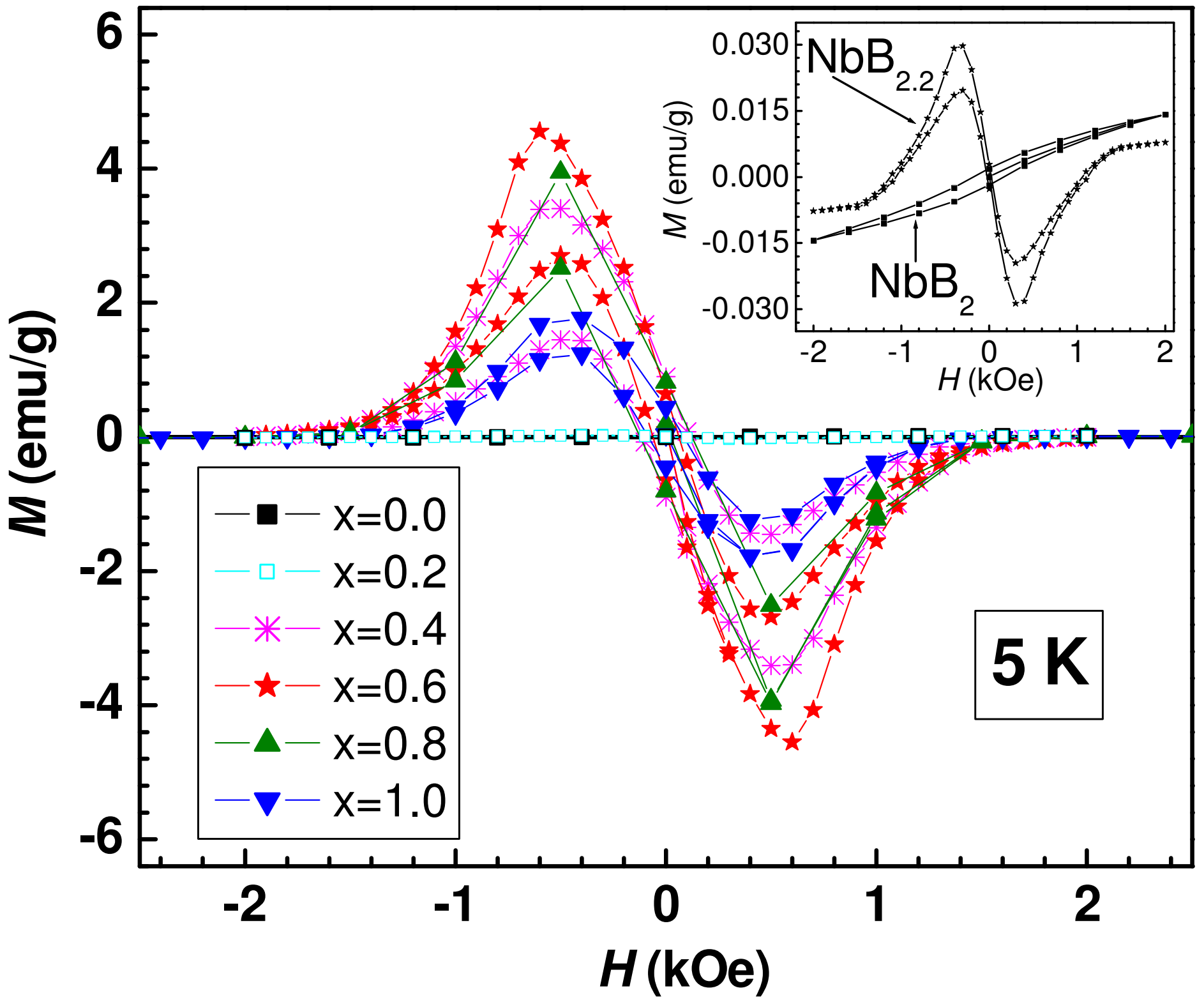

Fig.6(b)

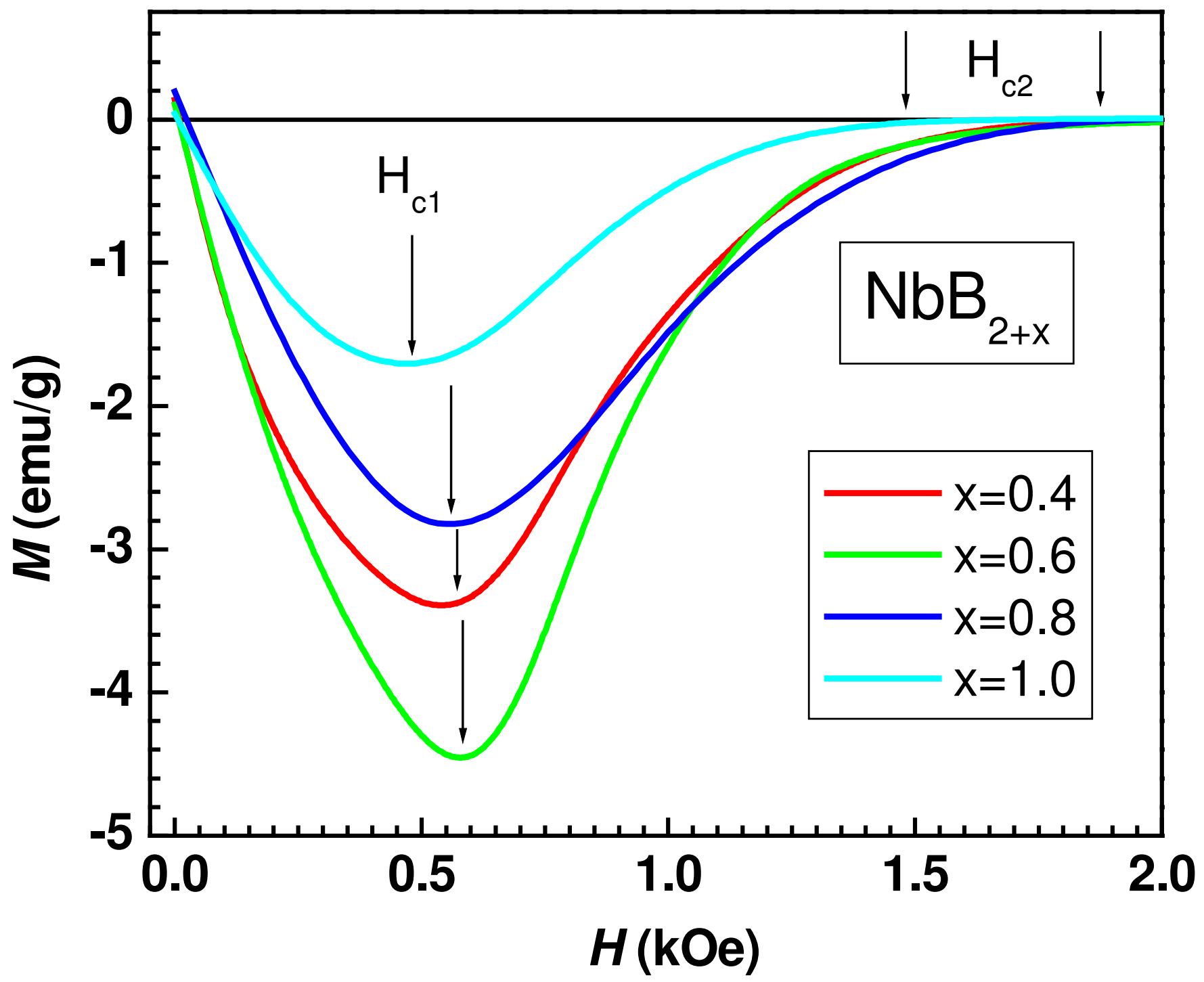